# Why is cheating wrong?

Mathieu Bouville
<mathieu.bouville@gmail.com>

Since cheating is obviously wrong, arguments against it (it provides an unfair advantage, it hinders learning) need only be mentioned in passing. But the argument of unfair advantage absurdly takes education to be essentially a race of all against all; moreover, it ignores that many cases of unfair (dis)advantages are widely accepted. That cheating can hamper learning does not mean that punishing cheating will necessarily favor learning, so that this argument does not obviously justify sanctioning cheaters.

**Keywords:** academic dishonesty, academic integrity, academic misconduct, education, ethics, homework, plagiarism

Asking why cheating is wrong may seem a silly question or a gratuitous provocation. Indeed, since "just about everyone agrees that cheating is bad and that we need to take steps to prevent it" (Kohn, 2007), no question seems warranted and no argument seems needed. Talk about cheating is then a matter of outrage: "students STOLE a password and then they used it to CHEAT" (Lingen, 2006), "plagiarism is WRONG no matter what the extent" (Parmley, 2000). A. J. Ayer would have said that a claim that cheating is wrong is just writing 'cheating' followed by some special exclamation mark — Lingen and Parmley prefer capitals. If asked why cheating is wrong, they may reply by using a larger font or boldface. Yet, typography is not a scientific method conducive to the truth: cheating is not wrong because it is capitalized.

A common view is that cheating is forbidden and cheaters break a rule. For instance, the focus of Burkill and Abbey (2004) on "regulations" and on "penalties" for "ignoring academic conventions" indicates that to them the main reasons for students to avoid cheating are obedience to rules and avoidance of penalties. However, Alfie Kohn (2007) draws attention to those "cases where what is regarded as cheating actually consists of a failure to abide by restrictions that may be arbitrary and difficult to defend". Breaking a rule is illegitimate only if the rule is legitimate. Either the rule has a rational justification and this rather than breaking a rule makes cheating wrong, or the rule is arbitrary and there is no reason to endorse it. In other words, cheating should be forbidden because it is wrong, not wrong because it is forbidden. Obviously, the wrongness of cheating should be an ethical not a bureaucratic question.

As Drake (1941) pointed out, cheating can be frustrating to the instructors, who may "interpret such behaviour as a direct affront to themselves." When Johnston (1991) found out that students had cheated she felt betrayed: "how could they do this to me?". While this may explain better than genuine arguments why teachers dislike cheating it does not show that cheating is wrong. It is interesting to note that this is generally not offered as an argument in articles looking at cheating in a 'cold' objective way but can be found in more personal papers, such as that of Johnston. This seems to acknowledge that this is both a real reaction of the 'victims' of cheaters and not perceived as a valid argument against cheating.

The research on cheating is empirical and focuses on quantification and correlations; yet finding out how many and which students cheat is of importance only if cheating itself is important. And cheating is important only if it is wrong. Since everything else depends on it, the question of the wrongness of cheating is the most important question. It is the object of this article.

# 1 CHEATING AND GRADES

It is a wide-spread error in issues of cheating to assume that cheating is obviously wrong (Bouville, 2007b; 2008b; 2008c). In particular, if one does not know why cheating is wrong one cannot set policies that would solve the problem. In this section, I will focus on the relationship between cheating and grades: cheaters receive undeservedly high grades and thus an unfair advantage over other students.

## 1.1 Grades as setting how good a student is

A fairly common view is that grades are the same as the value of the student. (I will use 'value' and 'worth' as shorthand for knowledge, talent, competence, etc.) For instance when their children get poor grades, parents start worrying. They may also demand that their child get better grades (e.g. threatening or promising). This may mean that the grade is an infallible evaluation of how good a student is, so that if grades are low it can only be because the student does not work enough. Another possibility is that grades actually define how good a student is, i.e. there is no concept of the worth of the student independent of grades. Since these views are seldom made explicit, what people actually believe is fuzzy. However, it is enough for us to see that many equate grades and how good the student is.

The main issue with this view is that if grade and value are the same then cheaters are good students, since they get good grades. When teachers give a bad course grade to a student they deem good, it must mean that their impression was mistaken since the grade *proved* that the student was not in actuality so good. This is the only possible conclusion if one takes grades to define how good students are (and this is naturally not a conclusion one can endorse). Plainly, if anything is to be said against cheating, one must recognize that grades are but a *proxy* for how good students are, an approximation of what they know, what they can do. It is thus possible for grades and worth to be different; every teacher has given grades that did not seem to correspond to what the students 'were worth' — some students are not good at taking tests, a student may have made a silly and costly mistake, etc. (also see Bouville, 2007a). Cheating is another source of discrepancy.

## 1.2 Cheating and the future success of the students

One should note that, since it is of the nature of grades to describe student performance, a grade that is a poor description is a poor quality grade. Such a grade is like a map of a city that does not actually represent the streets of this city; but if no one ever were to use this map, the problem would be a purely abstract one (and in particular it would not justify to blame and fire who drew the faulty map). Inaccurate grades matter only if someone somehow acts upon them. Otherwise I could just as well assign $-\pi$ as a grade, this would make no difference. Naturally grades are concretely used: grades are a proxy for what students know and can do, which is in turn used as a proxy for what students may be able to do in the future. In other words, grades are used as predictors of future success: high school grades are used for admission to universities; undergraduate grades for graduate school, law school grades to infer how good a lawyer the student will be, etc. In admissions, one looks at grades only in order to guess how well students may do in the future: how well they did in the past is not interesting *per se* — the past is important only inasmuch as it pertains to the future (if there were no correlation between grades in year n and academic success in year n+1 then the use of grades would plummet.) Therefore, any time grades do not correspond to how well students can be expected to perform a poor decision will be made. Such decisions can be called unfair since they may advantage less deserving students.

# 2 LIMITS OF THE ARGUMENT OF UNFAIR ADVANTAGE

## 2.1 The difficulty to use grades

Admission to universities is based on how students fare in high school, admission to graduate school on how good students are as undergraduates, etc. This means that one always uses someone else's criteria, even though they are not obviously relevant. High school is about learning (one may fare well with little talent if one has enough memory and is willing to spend time filling it) whereas higher education requires more understanding. In fact high school sometimes even penalizes the more gifted students, who may have difficulty following a fastidious (and to them useless) procedure or justifying something blatantly obvious, which high school teachers require from them for full credit (this is especially common in math). Likewise graduate students are supposed to invent new things, not just learn from others. Being able to digest what others found is handy but not sufficient to do good research — so one may be an excellent undergraduate but a poor PhD student. Using undergraduate results (beyond proof of adequate knowledge and understanding) to guess how good a PhD student will be means trying to infer apples from oranges. Why then would the quality of the oranges make any difference? The inference is flawed even in the absence of cheating.

There is another reason why a grade is of little help to estimate the future success of students. Hall *et al.* (1995) found that a deep approach to learning correlated negatively with SAT scores; students who merely learned by rote and who minimized their involvement or tried to get good grades without caring about what they learnt obtained higher scores than students who sought a deeper understanding of what they were taught. In other words, students who had good work habits and a sound mindset that would help them succeed in the long run received lower SAT scores — students most likely to succeed are treated as least desirable.

### 2.2 *Discipulus discipulo lupus*

There can be an unfair advantage only in those cases of direct competition between students. Entrance exams and other 'high stake' tests are an example. Homework is not. The argument of unfair advantage is thus limited in its scope. It will nevertheless be invoked in cases where it does not apply. How teachers see cheating is an interesting clue of how they see education. Taking cheating to be essentially a matter of unfair advantage means that education is one big race of all against all — *discipulus discipulo lupus*. For instance, the goal of homework is not the assignment of grades but rather to have students learn a lesson by putting it to practice; the main reason for grading homework is that some students may need such a carrot. Treating cheating on homework as essentially a matter of a student getting an unfair advantage means losing sight of what one is trying to accomplish. Not only is the focus generally on grades rather than on learning, grades rather than learning are seen as the issue in cases where grades are irrelevant.

### 2.3 Unfair advantage without cheating

It is common for teachers to knowingly give a student a grade that is evidently inadequate: for instance they commonly give a good student a bad grade, fully aware that the student deserves better... but the grade that came out of the exam is the grade that came out of the exam. And they do not mind doing so. When a grade is a poor assessment of the value of the student, it is the grade that wins (for instance, it is this faulty assessment that will be part of transcripts, not the actual worth of the student).[1] If one does not see a major problem with grades being decorrelated from how good students are then the fact that the grades of cheaters do not reflect their actual value should not be a problem either. In other words, there are cases in which no cheating is involved yet a grade is clearly a bad estimate of how good a student is, i.e. an unfair (dis)advantage. It is then unclear, if teachers are not bothered by such incongruities, why similar discrepancies would be problematic when due to cheating.

Picture a student who has an essay proofread by his parents or a personal tutor; the student did all the writing but received help that contributed to improving his work (e.g. that section is unclear, this book should be of help). He will get a better grade than a student of equal intelligence and talent who cannot receive or afford any such help. This is an unfair advantage but one would not call it cheating (whatever one thinks of the unfairness of the situation, the favored student did nothing wrong). If grades are used to decide who should be admitted to a top university, a smart and talented student, a student with a tutor, and a cheater will look the same even though the first is superior to the other two. Cheating and tutoring both create an unfair advantage.

---

[1] The reason why grades trump one's intuition of the value of students is probably that they are objective and thus superior to the subjective opinion of a teacher. But if grades claim that good students are bad, of what exactly are they an objective measure? Grading based on the number of points the student's name would get in Scrabble is objective as well; it is also completely silly (also see Bouville, 2007a). Saying that the objectivity of grades is their main quality means that what they actually measure is of secondary importance. "What grades offer is spurious precision" (Kohn, 1994). This, again, undermines the meaning of grades as measure of the value of the students.

## 3 Cheating and learning

### 3.1 Cheating undermines feedback

Passow *et al.* (2006) argue that "acts of academic dishonesty undermine the validity of measures of student learning". If teachers do not know that there is something the students do not understand (if they cheat it may seem that they understand) then it is impossible for them to know whether to accelerate or slow down, on what to focus, or how to re-design their lectures next year — in the long term, cheating hurts the students. It also prevents teachers from providing students with relevant feedback.

One should remark that this argument is more relevant to homework than to exams (especially final exams) because the latter are used more for grading or ranking and less for feedback, making cheating on homework worse than cheating on finals. Similarly, cheating on entrance exams would not be wrong at all since these are not meant to provide any feedback at all. In other words, this argument forces us to hold as worst the instances of cheating that would generally be seen as mildest. This is not surprising since feedback (either way) is not genuinely seen as of prime importance; that grades matter more is clearly reflected in the far greater importance given (by both students and teachers) to exams compared to homework. Finally, one should point out that if the only problem with cheating is merely that it hinders feedback then it is a very venial detail, and would not justify the outrage and dismissals one witnesses.

The applicability of this argument depends deeply on the actual practice of the teachers. In particular, it is not a universal truth that teachers use graded assignments for feedback to the students. Were it so, grades would be less ubiquitous and written comments far more numerous and extensive. Also, many teachers have taught the same class the same way for decades without ever changing their course based on the specifics of their class; so the fact that cheaters create noise on the feedback is irrelevant when this feedback is not taken into account anyway. In other words, not all instructors can claim that cheating interferes with their teaching. In fact, this argument could also be used against these instructors who do not provide students with useful information on how they are doing or who do not make use of the information they receive from their students.

### 3.2 Cheating undermines learning

A more important issue with cheating is that it can directly get in the way of learning. For instance, students who copy homework assignments instead of doing them themselves will not learn what they should. Likewise, having a book in one's lap does not have the same didactic impact as studying for an exam. For cheaters to be punished because cheating hinders learning, the following five conditions are necessary (the last of them —that sanctioning cheaters must actually have a positive consequence— will be addressed in the next section).

The assignment on which the student cheated must teach this student something worthwhile. Cheating must *actually* interfere with learning (that it may *a priori* is not enough). The best students may have little need for homework, some teachers assign work which has little pedagogical value, etc. Can students who would not learn anything by doing the homework copy it? It makes no sense to make certain students fail a class because they were so good that they did not need homework (also see Kohn, 2007). Moreover if the problem is that the teacher assigns work that does not contribute much to student learning one may wonder why the students are punished rather than the teacher.

Cheating on this assignment must hinder learning. One should remark that it may be the absence of cheating, rather than cheating, that hinders learning. For instance, Stephens (2005) found that "only 18 percent [of high school students] believed that 'working on an assignment with other students when the teacher asked for individual work' was cheating". This is because "students regarded this forbidden collaboration as furthering their knowledge and understanding, and therefore saw it as an act of learning rather than a form of cheating" (also see Kohn, 2007).

The sanction must not hamper learning more than cheating does. Sanctions can be quite dramatic (e.g. the University of Virginia expelled nearly fifty students for plagiarism in 2002). Expelling students so they do not fail later classes and eventually drop out is as meaningless as making suicide liable to death penalty because suicide is wrong. (Milder sanctions are not likewise logically flawed.)

Anything that hinders learning as much as cheating does must be sanctioned as much as cheating. Since, in terms of learning, not doing one's homework at all and copying it are on a par, the argument of hindrance to learning cannot justify treating cheaters more harshly. (Those who exhibit moral outrage at cheaters but not at students who study little, or more outrage at cheaters, are not reacting against a hindrance to learning. The same is true when one takes cheating on exams as far worse than cheating on homework.) Take three students. One is bright and learns nothing from doing a given assignment, another did not do the assignment and the last one copied it from a friend. One broke the "Thou shalt not cheat" commandment and two the commandment that "Thou shalt do the assignment", yet none of these three students got anything out of the assignment. From the viewpoint of learning there is no difference between them — why should one of them be sanctioned? (Hobbies, working for tuition money, etc. can adversely affect learning as well; yet one would not expel students just because they have a part-time job or a boyfriend.) [2]

## 4   SHOULD ONE REDUCE CHEATING?

One cannot but notice that arguments against cheating have something utopian about them: they would be forceful if grades always reflected how good students are and always enabled the prediction of their future success, and if teachers always used assignments to get information on the students as well as to provide them with relevant feedback on their performance. While systematic cheating hampers learning and distorts admission decisions, cheating once on an assignment of minor importance is clearly different in both respects. Cheating has a negative impact on education inasmuch as students cheat rather than study. But this justifies trying to curb cheating only if doing so actually has a positive impact on learning.

### 4.1   Curbing cheating for the sake of learning

Imagine an athlete who would compete in a marathon and who would take a shortcut in order to be more likely to win. Whatever one thinks of such behavior, one must recognize that this has a rationale: this is wrong but not absurd. On the other hand, a jogger who would take a shortcut would do something absurd rather than wrong: it makes no sense to take a shortcut when one runs to exercise, since there is no opponent over whom to get an advantage. Likewise it makes sense to cheat on a 'high stake' exam in order to get a better grade and (e.g.) be admitted to a better university. Regarding cheating on homework, on the other hand, either students indeed do something that makes no sense or grades are not the (only) reason for what they do.

Jensen *et al.* (2002) quote a high school student: "I'm a dedicated student, but when my history teacher bombards me with 50 questions due tomorrow or when a teacher gives me a fill-in-the-blanks worksheet on a night when I have swim practice, church, aerobics —and other homework— I'm going to copy from a friend!". Similarly, Cole and Kiss (2000) found that "students are most likely to cheat when they think their assignments are pointless, and less likely to cheat when they admire and respect their teachers and are excited about what they are learning" (also see Collier *et al.*, 2004; McCabe, 1997; McKeachie, 2002; Murdock *et al.*, 2004; Parameswaran, 2007). Grades are obviously not the only concern, otherwise better assignments and better teachers would have no impact on cheating (and students would not copy homework that will not be graded). Some students are not motivated by what they are taught and they copy the assignments so they do not waste their time on something of no interest to them (while at the same time getting good enough grades not to be in trouble). Jason Stephens and Heather Nicholson (in press) interviewed a student who is "simply not very interested in learning (or working hard at it) and he isn't much emotionally affected by his cheating, which he acknowledges is wrong." It is far from obvious that if this student stopped cheating he would study hard instead. Cutting the motivation to cheat will not automatically create a motivation to study: if cheating were reduced directly (by heavier proctoring or use of antiplagiarism software for instance) there is no reason to assume that students who just want to limit the impact of school on their lives would start studying more. That cheating has a negative impact on education does not mean that reducing cheating necessarily has a positive impact on education.

---

[2] One should also remark that grades (which one so dearly wants to protect from cheating) are bad for education as well. Ruth Butler (1988) found that students who received feedback in the form of grades did worse than those who received written comments but no grade. Butler and Nissan (1986) note that "grades may encourage an emphasis on quantitative aspects of learning, depress creativity, foster fear of failure, and undermine interest." According to Anderman and coworkers (1998), "students who reported cheating in science perceived their classrooms as being extrinsically focused and perceived their schools as being focused on performance and ability" — i.e. the emphasis on grades favors cheating.

When you can get something for free, you just take it; if it is no longer free, either you pay for it or you give it up. What would students do if they could no longer cheat (i.e. get good grades for free)? Some would indeed do the homework assignments and study for the exams (to keep their grade the same), but other students may not study more (to keep their amount of work the same). Even if grades were the only point of cheating, curbing cheating would not necessarily make students study more, i.e. it may not improve learning.

It is obvious that certain forms of cheating can get in the way of learning. It is obvious that education is about learning (even though one may challenge the importance given to grades in education, one cannot challenge the importance given to learning). So it is obvious that cheating can go against the very essence of education. On the other hand, the consequences in terms of policy, in particular regarding sanctions, are not obvious. If the problem with cheating is that it hinders education, there is no point to fight cheating if this does not positively affect education. While interference with learning should be the most obvious problem with cheating, it is not the one that most straightforwardly leads to efficient policies, i.e. policies that would protect learning from the threat of cheating.

### 4.2 Cheating in school correlates with cheating later on

Drake (1941) hoped that "the dishonesty so learned is specific and does not carry over to other activities." In fact, recent studies show that cheating as student correlates with cheating in one's professional life and with other misbehaviors (e.g. Blankenship and Whitley, 2000; Roig and Caso, 2005). One should first notice that this correlation does not seem to exist for all students: for instance, Mustaine and Tewksbury (2005) found that "cheating may be part of a larger problem behavior orientation for males but not females." Furthermore, this correlation is of importance only if lowering rates of cheating in school has lasting effects. McCabe *et al.* (1996) found "no significant differences between Code ($M$=0.95) and No-code College ($M$=1.00) alumnae/i on self-reported unethical behaviour ($t(281)$=.95, $p$=.61)" so that the hypothesis "individuals who experienced an honor code environment in college will self-report less unethical behaviour in the workplace than individuals who did not experience an honor code environment" is "not supported." In other words, even though honor codes decrease rates of cheating in university (Roth and McCabe, 1995), they do not seem to have a lasting effect. This is another example of a negative consequence of cheating that is not removed simply by directly removing the cheating.

## 5 CONCLUSION

Cheating is disliked to a great extent because it breaks a rule and because teachers take it as a personal offense. However, for cheating to be wrong one must justify the rule forbidding cheating. And the fact that the teacher dislikes what a student did does not necessarily mean that the student did something wrong. This article looked at two kinds of arguments against cheating: the unfair advantage it provides and the hindrance to learning it generates.

There can be an unfair advantage only when grades are used as proxies to estimate how students will fare in the future. The argument of unfair advantage is thus limited in its scope — mostly entrance exams and other 'high stake' tests. Taking cheating to be essentially a matter of unfair advantage means that education is essentially a fight of all against all or it means that cheating can occur (or matter) only in a few specific situations. The former idea is generally despised (even though what teachers in fact do may be consistent with it) and the latter would make cheating of limited importance and would thus challenge the usual handling of it. Another limitation of the argument is that cheating is not the only way to dissolve the link between grades and future success, so that one cannot say *a priori* that a given instance of cheating will get a student something undeserved in a way chance could not. For instance, it is common for teachers to knowingly give a good student a bad grade. But if they are not bothered by such incongruities, why would similar discrepancies be problematic when due to cheating?

It is obvious both that certain forms of cheating can get in the way of learning and that education is about learning. Cheating can thus plainly go against the very essence of education. On the other hand, this does not obviously entail that cheaters should systematically and harshly be punished. I noted that cheaters may be punished because cheating hinders learning only provided that cheating actually hindered learning (not just that it may have, *a priori*) and that the sanction does not make the situation worse (which would make it self-defeating). It would also make sense that anything that hampers learning (not just cheating) be likewise sanctioned. The main issue with this argument is that curbing cheating will not necessarily favor learning: those students who want to get passable grades with as little work as possible are unlikely to start studying hard because they can no longer cheat. And if curbing cheating does not have a major positive impact on learning then the fact that cheating hinders learning cannot justify sanctioning cheating.